\documentclass[doublecol]{epl2}
\usepackage{amssymb}
\usepackage{graphicx}
\usepackage{amsmath,amssymb}
\usepackage{xcolor}
\usepackage{nameref}
\title{Currents and current correlations in a topological superconducting nanowire beam splitter}
\author{Zhan Cao\inst{1} \and Tie-Feng Fang\inst{1} \and Qiao Chen\inst{2} \and Hong-Gang Luo\inst{1,3}}
\institute{
  \inst{1} Center of Interdisciplinary Studies $\&$ Key Laboratory for Magnetism and Magnetic Materials of the Ministry of Education, Lanzhou
University, Lanzhou 730000, China\\
  \inst{2} Department of Maths and Physics, Hunan Institute of Engineering, Xiangtan 411104, China\\
  \inst{3} Beijing Computational Science Research Center, Beijing 100084, China
}


\abstract{A beam splitter consisting of two normal leads coupled to one end of a topological superconducting nanowire via double quantum dot is investigated. In this geometry, the linear current cross-correlations at zero temperature change signs versus the overlap between the two Majorana bound states hosted by the nanowire. Under symmetric bias voltages the net current flowing through the nanowire is noiseless. These two features highlight the fermionic nature of such exotic Majorana excitations though they are based on the superconductivity. Moreover, there exists a unique local particle-hole symmetry inherited from the self-Hermitian property of Majorana bound states, which is apparently scarce in other systems. We show that such particular symmetry can be revealed through measuring the currents under complementary bias voltages.}

\begin{document}

\maketitle
\section{Introduction}\label{intr}
Over the last two decades, the hybrid multiterminal structures consisting of a BCS superconductor and two normal metal leads keep receiving extensive interest from both theoretical \cite{Torres1999,Recher2001,Melin2008,Freyn2010,Floser2013} and experimental \cite{Hofstetter2009,Herrmann2010,Schindele2012} communities. The generic physics contained in this versatile platform is the interplay between coherence effect in the normal leads and intrinsic coherence of the superconducting condensate. One of the most appealing advantages of these structures is to act as a Cooper pair beam splitter \cite{Choi2000,Lesovik2001}, which splits constituent spin-entangled electrons from the superconductor into the separate normal leads. This enables them as entanglement generating sources in quantum-information processing \cite{{Burkard2000,DiVincenzo2000}}. It also allows for the study of electronic Einstein-Podolsky-Rosen experiment based on current cross-correlations \cite{Chtchelkatchev2002,Bouchiat2003}. In general, the subgap transport occurring in these hybrid structures with two normal leads includes following elementary processes \cite{Melin2008,Freyn2010}: an electron emitted from one of the leads is reflected back as a hole, or is transmitted as an electron or a hole into the other lead. The former one is the traditional local Andreev reflection (AR) occurred in one of the contacts between the normal leads and the superconductor, while the latter two, termed as elastic cotunneling (EC) and crossed Andreev reflection (CAR), are nonlocal processes involving both two separated leads. At lowest order in the tunneling amplitudes between the normal leads and the superconductor, the two nonlocal processes are decoupled, leading to positive (negative) current cross-correlation for CAR (EC) process \cite{Bignon2004}. A simple interpretation of this fact is that CAR implies instantaneous currents of the same sign in both leads, while EC implies instantaneous currents of opposite signs. This positive current cross-correlation roughly signals the Bosonic nature of Cooper pairs to some extent\cite{Blanter2000}.

Recently, the interplay of AR, EC and CAR processes are predicted to lead remarkable transport behaviors when the BCS superconductor in these hybrid structures is replaced by a InAs or InSb  topological superconducting nanowire (TSNW) \cite{Law2009,Wu2012,Cao2012,Liu2014}, which is expected to support a pair of exotic end Majorana bound states (MBSs) \cite{Oreg2010,Mourik2012,Deng2012,Das2012,Churchill2013,Lee2014} that are promising candidate to realize nonlocal qubits free of local decoherence \cite{Alicea2011}. In particular, Zocher \textit{\textit{et al}.} studied the situation where the nanowire is sandwiched between two normal leads with inserted quantum dots (QDs) \cite{Zocher2013}. They found that the linear current cross-correlation at zero temperature is proportional to the overlap between the two end MBSs and thus vanishes for perfectly decoupled MBSs pair, and more particularly, is a positive definite quantity. This feature was also reported by Nilsson \textit{et al}. in a similar device but without QDs \cite{Nilsson2008}. However, such positive current cross-correlation looks to be a consequence of boson-like excitation, which is inconsistent with the fact that MBSs are fermionic excitations that obey canonical anti-commutation relation.

\begin{figure}[top]
\begin{center}
\includegraphics[width=0.9\columnwidth]{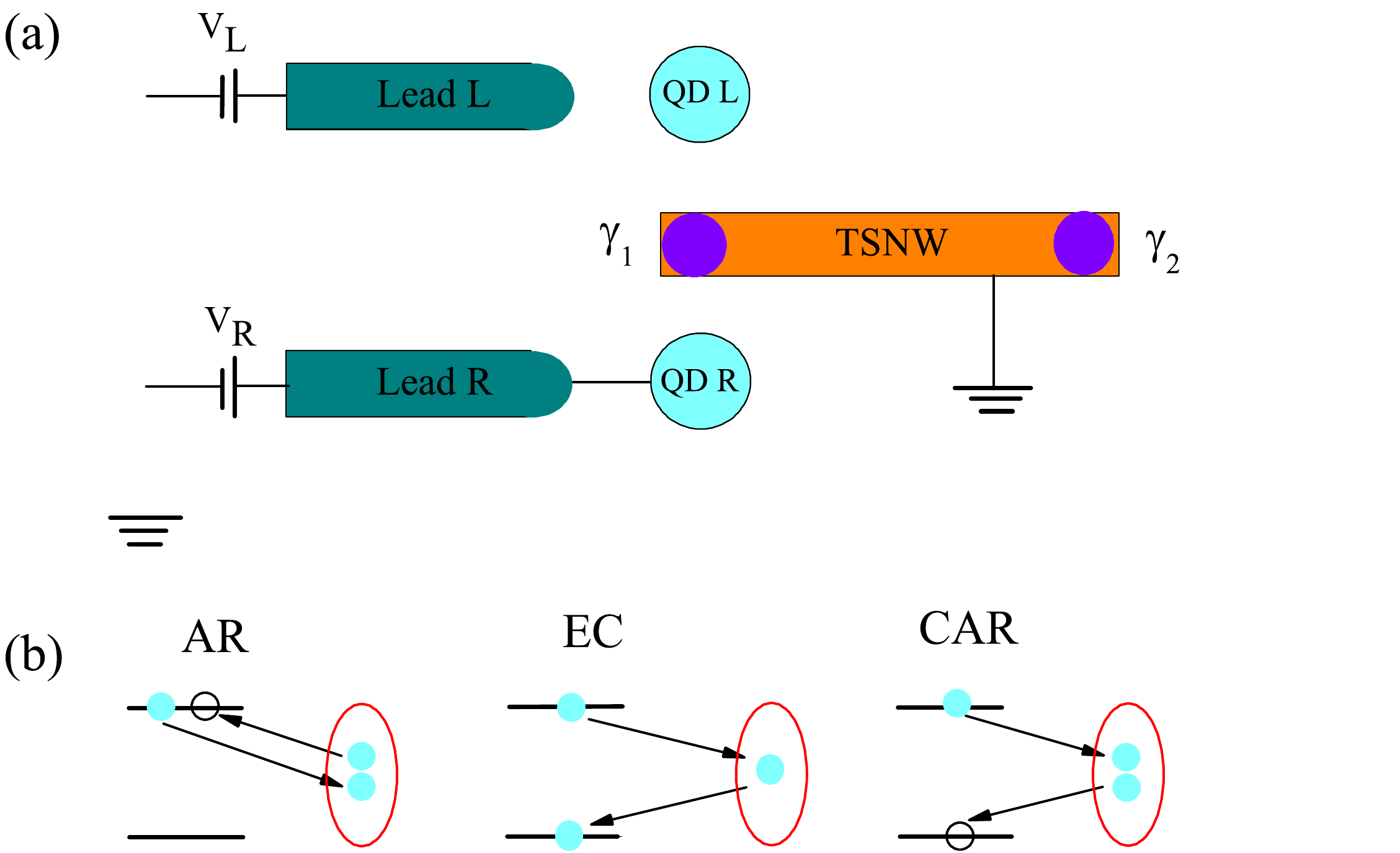}
\caption{(Color online) (a) Schematic view of our device. Two QDs are attached to separate normal leads and coupled to the nearest MBS hosted by a grounded TSNW. (b) Three elementary transport processes: an electron emitted from one normal lead is reflected back as a hole (AR), transmitted as an electron (EC) or a hole (CAR) into the other lead.}\label{fig1}
\end{center}
\end{figure}

In this paper, we investigate a beam splitter consisting of two normal leads coupled to one of the MBSs hosted at the ends of a TSNW through double noninteracting QDs, as shown in fig.\,\ref{fig1}(a). The inserted QDs provide a controllable way to achieve deep insight into the interplay of the elementary transport processes in virtue of the high tunability of QDs \cite{Chevallier2011,Rech2012}. In this setup we find that, under appropriate bias voltages and dot levels, the current cross-correlations expected to be positive (negative) in conventional BCS Cooper pair splitters are indeed negative (positive) in our device for weakly overlapped MBSs, but change signs towards strongly overlapped MBSs. Additionally, under the symmetric bias voltages the net current flowing through the nanowire is found to be noiseless. These two features on current correlations highlight the fermionic nature of such exotic MBSs though they are based on the superconductivity. Moreover, there exists a unique local particle-hole (p-h) symmetry in this system due to the self-Hermitian property of MBSs. This local p-h symmetry guarantees the transmission coefficient of charges emitting from one lead to the other lead by the EC process is identical to that by the CAR process through the reversed dot level, which can be revealed by measuring the currents under complementary bias voltages.

\section{Model}\label{model}
We consider two noninteracting single-level QDs connected to conducting electron reservoirs and to one end of a grounded TSNW, as shown schematically in fig.\,\ref{fig1}(a). Each QD is attached to a separate normal lead with controllable chemical potential $\mu_\alpha=-eV_\alpha$ ($\alpha=L,R$). Generally, a strong magnetic field is needed to drive the nanowire into the topological nontrivial phase \cite{Mourik2012,Deng2012,Das2012,Churchill2013,Lee2014}. We assume that the magnetic field is applied on the whole setup, and the bias voltage window is smaller than the large Zeeman splitting. Therefore, our considerations are limited to the case of spinless electrons.

The system under consideration can be described by the extended Anderson Hamiltonian of the general form
\begin{equation}
H=H_{\textrm{leads}}+H_{\textrm{central}}+H_{\textrm{tunnel}}. \label{eq1}
\end{equation}
The first term, $H_{\textrm{leads}}$, describes the two noninteracting normal leads, $H_{\textrm{leads}}=H_L+H_R$, with $H_\alpha=\sum_k \varepsilon_{k\alpha}c_{k\alpha}^\dag c_{k\alpha}$ being the Hamiltonian of lead $\alpha$. Here, $c_{k\alpha}^\dag(c_{k\alpha})$ is the creation (annihilation) operator of an electron with the wave vector $k$ in the lead $\alpha$, whereas $\varepsilon_{k\alpha}$ denotes the corresponding single-particle energy. The second term of Hamiltonian ({\ref{eq1}) describes the central region of the system,
\begin{equation}
H_{\textrm{central}}=\sum_{\alpha}\varepsilon_{\alpha}d_{\alpha}^{\dag}d_{\alpha}+i\varepsilon_{M}\gamma_{1}\gamma_{2} +\sum_{\alpha}\lambda_{\alpha}(d^\dag_\alpha -d_\alpha)\gamma_{1}, \label{eq2}
\end{equation}
where $\varepsilon_\alpha$ is the discrete level of the $\alpha$th QD. Since we are interested in the low-energy physics, we assume the induced superconducting gap in the nanowire is the largest energy scale and the applied bias voltage is controlled within the gap, thus the topological phase in the nanowire can be described by a low-energy effective model in which the two zero-energy MBSs hosted at two ends of the nanowire are represented by the Majorana operators $\gamma_i$ ($i=1,2$) \cite{Flensberg2010,Leijnse2011,Golub2011,Liu2011,Lopez2014,Cheng2014}, obeying the Clifford algebra $\{\gamma_i,\gamma_j\}=\delta_{ij}$ and the self-Hermitian property $\gamma_i=\gamma_i^\dag$. Moreover, for a finite-length nanowire, the two MBSs would couple to each other by a nonzero overlap energy $\varepsilon_{M}\sim \textrm{exp}(-l/\xi)$ with $l$ being the length of the nanowire and $\xi$ the superconducting coherence length. The last term in Hamiltonian (\ref{eq2}) corresponds to the tunnel coupling between the QDs and the nearest MBS $\gamma_1$. The tunneling between the QDs and the normal leads reads
\begin{equation}
H_{\textrm{tunnel}}=\sum_{k\alpha}( V_{\alpha}d_{\alpha}^{\dag}c_{k\alpha}+\textrm{H.c.}),\label{eq3}
\end{equation}
with $V_\alpha$ being the tunneling matrix elements. An electron and/or hole transfer between the QD and corresponding lead is described by an effective tunneling rate $\Gamma_\alpha=\pi\rho|V_\alpha|^2$, where $\rho$ is the density of states of the lead electrons.

It is seen that there exists a unique local p-h symmetry in this system, that is, after explicit local p-h transformation on operators concerning one of the QDs (say the right QD) as $d_R\rightarrow d_R^\dag$, $c_{kR}\rightarrow c_{kR}^\dag$, $\gamma_i\rightarrow\gamma_i^\dag$, $V_R\rightarrow-V_R$, $\varepsilon_{kR}\rightarrow-\varepsilon_{kR}$, and $\varepsilon_R\rightarrow-\varepsilon_R$, when the terms involving the left QD keep fixed, the total Hamiltonian is invariant due to the nature $\gamma_i^\dag=\gamma_i$ of MBSs. As we shall see, this local p-h symmetry can be revealed through measuring the currents under complementary bias voltages.
\section{Transport formulae}\label{formulae}
Following the standard nonequilibrium Green's function formalism \cite{Haug2008}, the current flowing through the $\alpha$th lead can be calculated from evolution of the total number operator of electrons in that lead, $N_\alpha=\sum_k c_{k\alpha}^\dag c_{k\alpha}$. For superconducting system, it is convenient to define the GFs in the Nambu space in a matrix form as $G_{\alpha\beta}=\left(
\begin{array}
[c]{cc}
\langle\langle d_{\alpha},d_{\beta}^{\dag}\rangle\rangle  & \langle\langle d_{\alpha}  ,d_{\beta}\rangle\rangle \\
\langle\langle d_{\alpha}^{\dag},d_{\beta}^{\dag}\rangle\rangle  & \langle\langle d_{\alpha}^{\dag},d_{\beta}\rangle\rangle
\end{array}
\right)$, with $\alpha=\beta$ ($\alpha\ne\beta$) representing the local (nonlocal) GFs between the same (different) QD. Accordingly, we denote the two elements in the upper row of $G_{\alpha\beta}$ as $G_{\alpha\beta;ee}$ and $G_{\alpha\beta;eh}$ while those in the lower row as $G_{\alpha\beta;he}$ and $G_{\alpha\beta;hh}$. Within such representation, the current flowing through lead $\alpha$ reads
\begin{equation}
I_{\alpha}=\frac{1}{2h}\int d\omega \textrm{Tr}\{\sigma[G_{\alpha\alpha}\Sigma_{\alpha}]^{<}-\sigma[\Sigma_{\alpha}G_{\alpha\alpha}]^{<}\},\label{eq4}
\end{equation}
where $\sigma=\left(
\begin{array}
[c]{cc}%
q^e & 0\\
0 &q^h
\end{array}
\right)$ accounts for the different type charge carried by the electrons $(q^e=e)$ and the holes $(q^h=-e)$. The correlation function between current fluctuations in lead $\alpha$ and $\beta$ is defined as $S_{\alpha\beta}(t)=\langle \{\delta I_{\alpha}(t)  ,\delta I_{\beta}(  0)\}$. Its Fourier transform gives the current-noise power spectrum as $S_{\alpha\beta}(\omega) =\int dte^{i\omega t}S_{\alpha\beta}(t)$. The general form of the power spectrum in the zero-frequency limit reads
\begin{eqnarray}
&S_{\alpha\beta}\equiv S_{\alpha\beta}(0)=\frac{1}{h}\int d\omega\text{Tr}\{  \delta_{\alpha\beta}(G_{\alpha\alpha}^{<}\sigma\Sigma_{\alpha}^{>}\sigma+\sigma\Sigma_{\alpha}^{<}\sigma G_{\alpha\alpha}^{>}) \notag \\
&+G_{\alpha\beta}^{<}\sigma[ \Sigma_{\beta}G_{\beta\alpha}\Sigma_{\alpha}]  ^{>}\sigma-\sigma[  \Sigma_{\alpha}G_{\alpha\beta}]  ^{<}\sigma[  \Sigma_{\beta}G_{\beta\alpha}]  ^{>} \notag \\
&-[  G_{\alpha\beta}\Sigma_{\beta}]  ^{<}\sigma[G_{\beta\alpha}\Sigma_{\alpha}]  ^{>}\sigma+\sigma[\Sigma_{\alpha}G_{\alpha\beta}\Sigma_{\beta}]  ^{<}\sigma G_{\beta\alpha}^{>}\}. \label{eq5}
\end{eqnarray}
The enclosed GFs in eq.\,(\ref{eq4}) and eq.\,(\ref{eq5}) can be expanded using the so-called analytic continuation rules \cite{Haug2008}. Noting that such an useful identity $(G^r_{\alpha\alpha}-G^a_{\alpha\alpha})_{mm}=\sum_{\beta=L,R}\sum_{n=e,h}G^r_{\alpha\beta;mn}(\Sigma^r_{\beta}-\Sigma^a_{\beta})_{nn}G^a_{\beta\alpha;nm}$ and the Keldysh equation $G^{<}_{\alpha\alpha;mm}=\sum_{\beta=L,R}\sum_{n=e,h}G^r_{\alpha\beta;mn}\Sigma^{<}_{\beta;nn} G^a_{\beta\alpha;nm}$. In the noninteracting case, the self-energies results from tunnel coupling between the QDs and the normal leads have the form
$\Sigma^{r,a}_\alpha=\mp i\Gamma_\alpha\left(
\begin{array}
[c]{cc}%
1 & 0\\
0 &1
\end{array}
\right)$ and $\Sigma^{<}_\alpha=2i\Gamma_\alpha\left(
\begin{array}
[c]{cc}%
f_\alpha^e & 0\\
0 &f_\alpha^h
\end{array}
\right)$.
After some algebra we arrive at the final current formula consisting of three elementary transport processes, shown schematically in fig.\,\ref{fig1}(b), as follow
\begin{eqnarray}
&I_\alpha=\sum_{m=e,h}\frac{q^m}{2h}\int d\omega[\underbrace{\mathcal{T}_{\alpha\alpha}^{m\bar m}(f_{\alpha}^{m}-f_{\alpha}^{\bar m})}_{\textrm{AR}}\nonumber\\
&+\underbrace{\mathcal{T}_{\alpha\bar\alpha}^{mm}( f_{\alpha}^{m}-f_{\bar\alpha}^{m})}_{\textrm{EC}}+\underbrace{\mathcal{T}_{\alpha\bar\alpha}^{m\bar m}(f_{\alpha}^{m}-f_{\bar\alpha}^{\bar m})}_{\textrm{CAR}}],\label{eq6}
\end{eqnarray}
with corresponding transmission coefficients
\begin{equation}
\mathcal{T}_{\alpha\beta}^{mn}(\omega)=4\Gamma_{\alpha}\Gamma_{\beta}\vert G_{\alpha\beta;mn}^{r}(\omega)\vert ^{2}. \label{eq7}
\end{equation}
Here $\bar\alpha$ and $\bar m$ denote the index opposite to $\alpha$ and $m$. $f_{\alpha}^{e,h}(\omega)=[\exp(\frac{\omega\mp eV_{\alpha}}{k_{B}T})+1]^{-1}$ are the Fermi distributions of the electron and hole states in the $\alpha$th normal lead. Note that due to the large gap assumption in our low-energy effective model, the quasi-particle tunneling between the normal leads and the TSNW is forbidden in this subgap regime. Differently from the situation for current, it is, in general, impossible to combine all the terms in current-noise power spectrum to module squares of GFs \cite{Freyn2010}, as well as of scattering matrixes \cite{Buttiker1990,Martin1992,Floser2013}. The current formula eq.\,(\ref{eq6}) reduces to instructive form when applying specific bias voltage configuration,
\begin{equation}
I_\alpha=\frac{e}{2h}\sum_{m=e,h}\int d\omega(\mathcal{T}_{\alpha\alpha}^{m\bar m}+\mathcal{T}_{\alpha\bar\alpha}^{m\bar m})(f_{\alpha}^{e}-f_{\alpha}^{h}), \label{eq8}
\end{equation}
for the symmetric bias voltages (SBV) $V_L=V_R=V$ where the EC process is completely blockaded, and
\begin{equation}
I_\alpha=\frac{e}{2h}\sum_{m=e,h}\int d\omega(\mathcal{T}_{\alpha\alpha}^{m\bar m}+\mathcal{T}_{\alpha\bar\alpha}^{mm})(f_{\alpha}^{e}-f_{\alpha}^{h}), \label{eq9}
\end{equation}
for the antisymmetric bias voltages (ASBV) $V_L=-V_R=V$ where instead the CAR process disappears.

In the noninteracting case the exact GFs can be accessed with analytical expressions, and so do the transmission coefficients. The matrix elements of local and nonlocal GFs can be obtained through the Dyson equation $G_{\alpha\beta;mn}=\delta_{\alpha\beta}\delta_{mn}g_{\alpha\alpha;mm}+g_{\alpha\alpha;mm}V_{\alpha\gamma;m1}G_{\gamma\gamma;11}V_{\gamma\beta;1n}g_{\beta\beta;nn}$,
where we have employed two facts that (i) the bare GFs for QD $\alpha$ decoupled from the MBSs is in a diagonal form as $g^{r,a}_{\alpha\alpha}(\omega)=\left(
\begin{array}
[c]{cc}%
\frac{1}{\omega-\varepsilon_{\alpha}\pm i\Gamma_\alpha} & 0\\
0 &\frac{1}{\omega+\varepsilon_{\alpha}\pm i\Gamma_\alpha}
\end{array}
\right)$, and more importantly (ii) the particular hopping matrix between QD $\alpha$ and the MBSs is $V_{\alpha\gamma}=\left(
\begin{array}
[c]{cc}%
\lambda_{\alpha} &0\\
-\lambda_{\alpha} &0
\end{array}
\right)$ and $V_{\gamma\alpha}=V_{\alpha\gamma}^{\dag}$, which accounts for the self-Hermitian property of MBSs. The dressed GFs of MBSs $G_{\gamma\gamma}$ can be calculated through $G_{\gamma\gamma}^{-1}=g_{\gamma\gamma}^{-1}-\sum_{\alpha=L,R}V_{\gamma\alpha}g_{\alpha\alpha}V_{\alpha\gamma}=\left(
\begin{array}
[c]{cc}
\omega-\Sigma_\gamma(\omega) & -i\varepsilon_{M}\\
i\varepsilon_{M} & \omega
\end{array}
\right)$ with the self-energy $\Sigma_\gamma^{r,a}(\omega)=\sum_{\alpha=L,R}(\frac{\lambda_{\alpha}^{2}}{\omega-\varepsilon_\alpha\pm i\Gamma_\alpha}+\frac{\lambda_{\alpha}^{2}}{\omega+\varepsilon_\alpha\pm i\Gamma_\alpha})$. Combining these terms we immediately obtain the matrix elements of GFs as
\begin{equation}
G_{\alpha\beta;mn}=\delta_{\alpha\beta}\delta_{mn}g_{\alpha\alpha;mm}+\frac{V_{\alpha\gamma;m1}g_{\alpha\alpha;mm}V_{\gamma\beta;1n}g_{\beta\beta;nn}}
{\omega-\Sigma_\gamma(\omega)-\varepsilon_{M}^{2}/\omega}.\label{eq10}
\end{equation}
\section{Results and discussions}\label{dis}
We investigate the current and current correlations in such topological superconducting beam splitter under complementary bias voltages. In what follows, we consider the case $\Gamma_L=\Gamma_R\equiv\Gamma$ and take $\Gamma$ as the energy unit. Furthermore, to be more realistic we adopt asymmetric couplings between the two QDs and the MBS $\gamma_1$ in calculations. All results are obtained in the zero temperature limit.

\begin{figure}[top]
\begin{center}
\includegraphics[width=0.85\columnwidth]{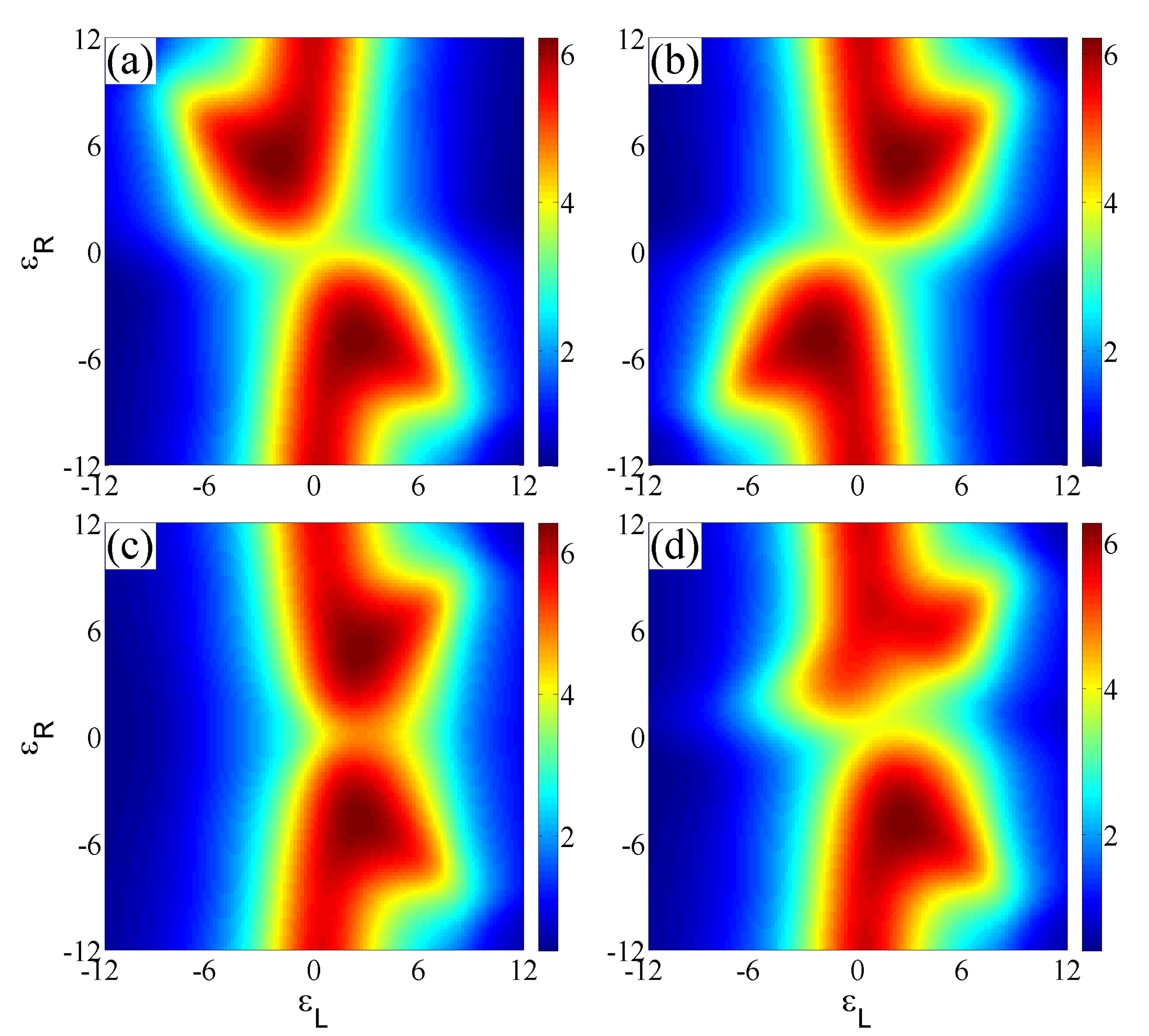}
\caption{(Color online) Density plots of the current $I_L$ (in unit of $e\Gamma/h$) flowing through the left lead as functions of the dot levels under (a) symmetric bias voltage $V_L=V_R=10$, (b) antisymmetric bias voltage $V_L=-V_R=10$, (c) $V_L=10$, $V_R=0$ and (d) $V_L=10$ , $V_R=4$. Other parameters are: $\varepsilon_M=0$, $\lambda_L=4$ and $\lambda_R=3$. Note the symmetries of currents displayed in (a)-(c) persist for nonzero $\varepsilon_M$ since the inherent local p-h symmetry is independent of the overlap between MBSs. }  \label{fig2}
\end{center}
\end{figure}

In fig.\,{\ref{fig2}} we show the current $I_L$ flowing through the left lead as functions of the dot levels under complementary bias voltages. In fig.\,\ref{fig2}(a) and fig.\,\ref{fig2}(b), the currents with dot levels $(\varepsilon_L,\varepsilon_R)$ are identical to those with $(-\varepsilon_L,-\varepsilon_R)$. This symmetry is attributed to the combination of the global p-h symmetry of the system and the particular bias voltage configuration. To be more precise, after the global p-h transformation the Hamiltonian is invariant but the roles of electrons and holes are exchanged. As a consequence, the transmission coefficients submit to the identity that $\mathcal{T}_{\alpha\beta}^{mn}(\varepsilon_L,\varepsilon_R)=\mathcal{T}_{\alpha\beta}^{\bar m\bar n}(-\varepsilon_L,-\varepsilon_R)$, which can also be readily inspected by eq.\,(\ref{eq7}) and (\ref{eq10}). This identity guarantees the invariance of currents addressed by eq.\,(\ref{eq8}) and eq.\,(\ref{eq9}) after reversing both dot levels. Another interesting symmetry of the currents is that the currents with dot levels $(\varepsilon_L,\varepsilon_R)$ in fig.\,\ref{fig2}(a) are identical to those with $(\varepsilon_L,-\varepsilon_R)$ in fig.\,\ref{fig2}(b). However, as we noted above, the involved elementary transport processes under SBV and ASBV are essentially different. This symmetry of currents is due to the nontrivial local p-h symmetry of the system, which indicates that $\mathcal{T}_{LR}^{m\bar m}(\varepsilon_L,\varepsilon_R)=\mathcal{T}_{LR}^{mm}(\varepsilon_L,-\varepsilon_R)$. Therefore, comparing $I_L^{\textrm{S}}$ and $I_L^{\textrm{AS}}$, it renders the contribution to the total current from the AR and CAR processes under SBV with dot levels $(\varepsilon_L,\varepsilon_R)$ being the same with that from the AR and EC processes under ASBV with dot levels $(\varepsilon_L,-\varepsilon_R)$. In fig.\,\ref{fig2}(c), the chemical potential of right lead is set to be aline with the Fermi level. In this case, the currents with dot levels $(\varepsilon_L,\varepsilon_R)$ are identical to those with $(\varepsilon_L,-\varepsilon_R)$, which is also introduced by the local p-h symmetry of Hamiltonian (\ref{eq1}) and the particular bias scheme. However, as shown in fig.\,\ref{fig2}(d), the symmetries of currents can not be uncovered under general asymmetric bias voltages. It is worth noticing that these intriguing symmetries of currents originated from the local p-h symmetry of the system is apparently not presented in other non-MBSs systems.

The maximum currents under SBV are along the line $\varepsilon_L=-\varepsilon_R$ [fig.\,\ref{fig2}(a)] while under ASBV are along the line $\varepsilon_L=\varepsilon_R$ [fig.\,\ref{fig2}(b)]. In the former case, where the EC process is prohibited, an electron transferred through the left dot level $\varepsilon_L$ will be resonant transmitted as a hole to the other lead when the right dot level equals $-\varepsilon_L$ since the Cooper pairs are residing near the Fermi level. On the contrary, in the latter case the EC process is allowed while the CAR process is instead prohibited. An electron injected through the left QD is to be resonant transmitted as an electron when $\varepsilon_L=\varepsilon_R$. Besides, the currents in each plots in fig.\,\ref{fig2} also exhibit resonant peaks centered around $\varepsilon_L=0$, which are results of the local AR process occurring between the left lead and the TSNW.

\begin{figure}[hc]
\begin{center}
\includegraphics[width=0.75\columnwidth]{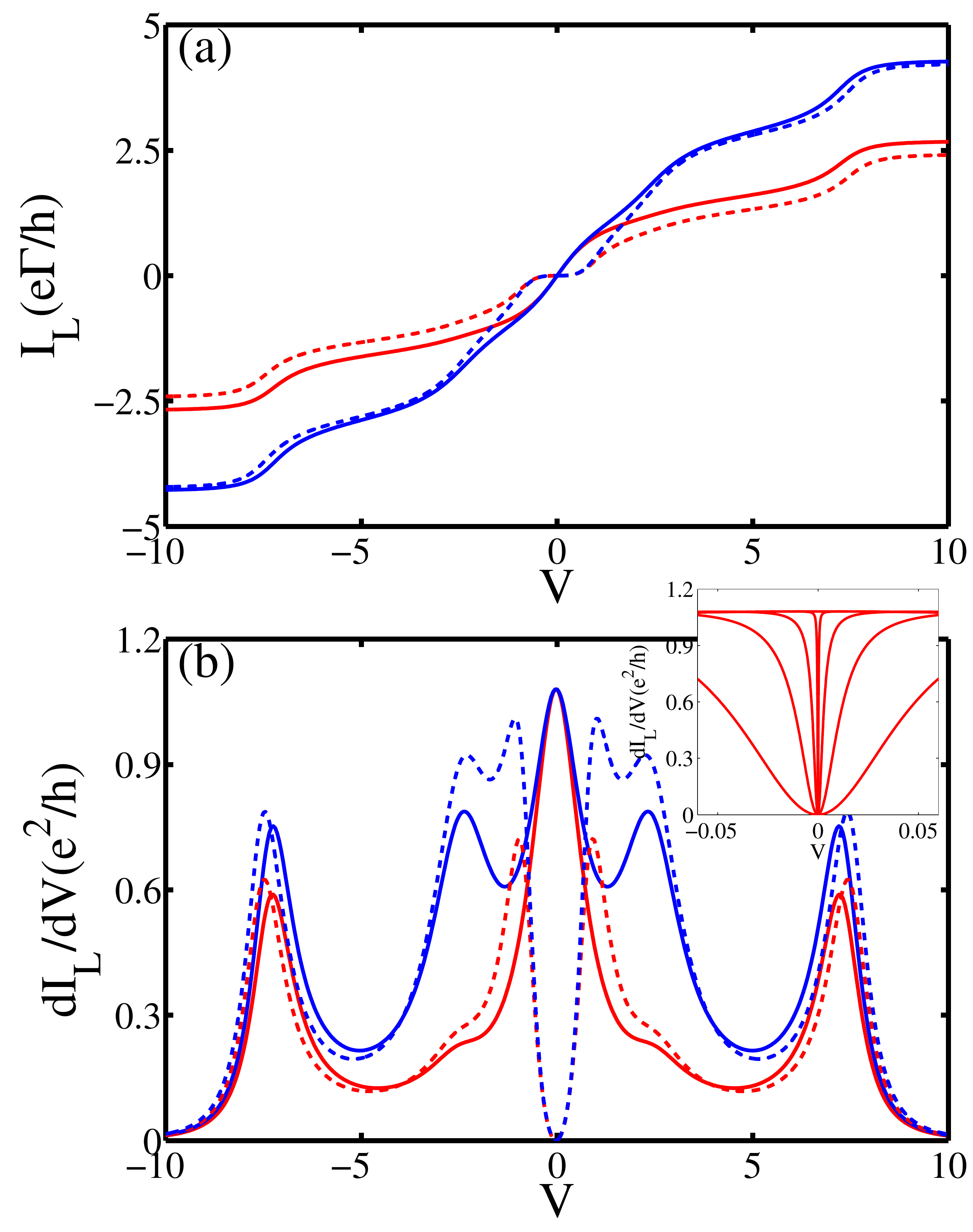}
\caption{(Color online) (a) The $I_L$ as a function of bias voltage under SBV (red) and ASBV (blue). The solid and dashed lines denote $\varepsilon_M=0$ and $\varepsilon_M=2$, respectively. (b) The corresponding differential conductances $dI_L/dV$. Inset: Evolution of the linear response conductance under SBV with increased overlap as $\varepsilon_M=0, 0.03, 0.1, 0.2, 0.4$. The parameters are $\varepsilon_L=4$, $\varepsilon_R=-2$, $\lambda_L=4$ and $\lambda_R=2$.} \label{figIV}
\end{center}
\end{figure}

The $I$-$V$ characteristics under SBV and ASBV with and without overlap between the two end MBSs are displayed in fig.\,\ref{figIV}. In the low-bias regime, the currents increase linearly as the voltage, as shown in fig.\,\ref{figIV}(a). The slopes corresponding to strong overlap $\varepsilon_M\sim\lambda_L, \lambda_R$ are much smaller than those to zero overlap, which are clearly shown by the differential conductances in fig.\,\ref{figIV}(b). For zero overlap, there are distinct zero-bias conductance peaks arising from the resonant EC, AR and CAR processes via the zero-energy MBS $\gamma_1$ with different weights depending on the specific bias voltages and dot levels. However, as the overlap becomes nonzero, the resonant peak splits into double peaks around the Fermi level symmetrically with a central dip [inset of fig.\,\ref{figIV}(b)]. In the nonlinear regime, all the differential conductances peaks under SB (red lines) are higher than those under ASB (blue lines), which is due to the dot levels we adopt have opposite signs that favors the CAR process. If the right dot level is reversed, the $I$-$V$ curves under SBV and ASBV will exchange to each other, as we discussed in fig.\,\ref{fig2}. Around the zero-bias, the differential conductances under SBV and ASBV are nearly identical [fig.\,\ref{figIV}(b)]. This feature is ascribed to the zero-energy transmission coefficient of EC process is identical to that of CAR process as
\begin{equation}
\mathcal{T}_{\alpha\bar\alpha}^{mm}(0)=\mathcal{T}_{\alpha\bar\alpha}^{m\bar m}(0)=\sqrt{\mathcal{T}_{\alpha\alpha}^{m\bar m}(0)\mathcal{T}_{\bar\alpha\bar\alpha}^{m\bar m}(0)},\label{eq11}
\end{equation}
with $\sqrt{\mathcal{T}_{\alpha\alpha}^{m\bar m}(0)}=\frac{\Gamma_\alpha \lambda_\alpha^2(\varepsilon_{\bar\alpha}^2+\Gamma_{\bar\alpha}^2)}{\Gamma_L \lambda_L^{2}(\varepsilon_R^{2}+\Gamma_R^{2})+\Gamma_R \lambda_R^{2}(  \varepsilon_L^{2}+\Gamma_L^{2})}$ ($\varepsilon_M=0$). Note this equivalence between zero-energy transmission coefficients does not require the sign reversal of either dot level. We would like to mention two nontrivial features on equation (\ref{eq11}). (i) If $\lambda_L=\lambda_R$, the zero-energy transmission coefficients are independent of the QDs-MBS coupling strengthes since $\lambda_L$ and $\lambda_R$ cancel out each other. (ii) At the high symmetric points $\Lambda_\pm$ ($\Gamma_L=\Gamma_R\equiv\Gamma$, $\varepsilon_L=\pm\varepsilon_R$, $\lambda_L=\lambda_R$), all three transmission coefficients turn out to be $1/4$, which is independent of the particular dot levels as well as the QDs-MBS coupling strengthes.

In fig.\,\ref{figS12} the current cross-correlations under SBV and ASBV are calculated. Similar to the currents in fig.\,\ref{fig2}, the cross-correlations under SBV with dot levels $\varepsilon_L=-\varepsilon_R$ are identical to those under ASBV with dot levels $\varepsilon_L=\varepsilon_R$ except a negative sign, which also resultes from the local p-h symmetry of this system. As the voltage increases, more and more tunneling channels are opened but the correlations vary nonmonotonically with $V$, depending on the cooperative or competitive relationship between these tunneling channels through which charges transferred. The cross-correlation between same type charges is always negative while between opposite type charges is always positive \cite{Datta1996}. As a result, the total cross-correlation can be either positive or negative up to the relative strengths of the above two terms. In the linear regime, where only the MBSs related low-energy channel is activated, the cross-correlation under SBV is negative for small overlap $\varepsilon_M$ (see red solid and dashed lines). This is a counterintuitive feature since the dot levels are chosen to be $\varepsilon_L=-\varepsilon_R$ to evoke resonant CAR process that likely to contribute positive cross-correlation \cite{Bignon2004}. Nevertheless, as the overlap $\varepsilon_M$ is increased the linear cross-correlation becomes positive as one may expected in the conventional Cooper pair beam splitters (see red dotted line). Similarly, the linear cross-correlation under ASBV with resonance condition $\varepsilon_L=\varepsilon_R$ for EC process is unexpectedly positive for weak overlap but turns to be positive for strong overlap (see blue lines). These seemingly anomalous signs of linear cross-correlations are appreciable since the MBSs are predicted to be fermionic excitations, however, the strongly overlapped MBSs are akin to standard Andreev bound states. We note these sign reversals of cross-correlations versus the overlap are absent when the TSNW is sandwiched between two normal leads with or without inserted QDs \cite{Nilsson2008,Zocher2013}.

\begin{figure}[top]
\begin{center}
\includegraphics[width=0.8\columnwidth]{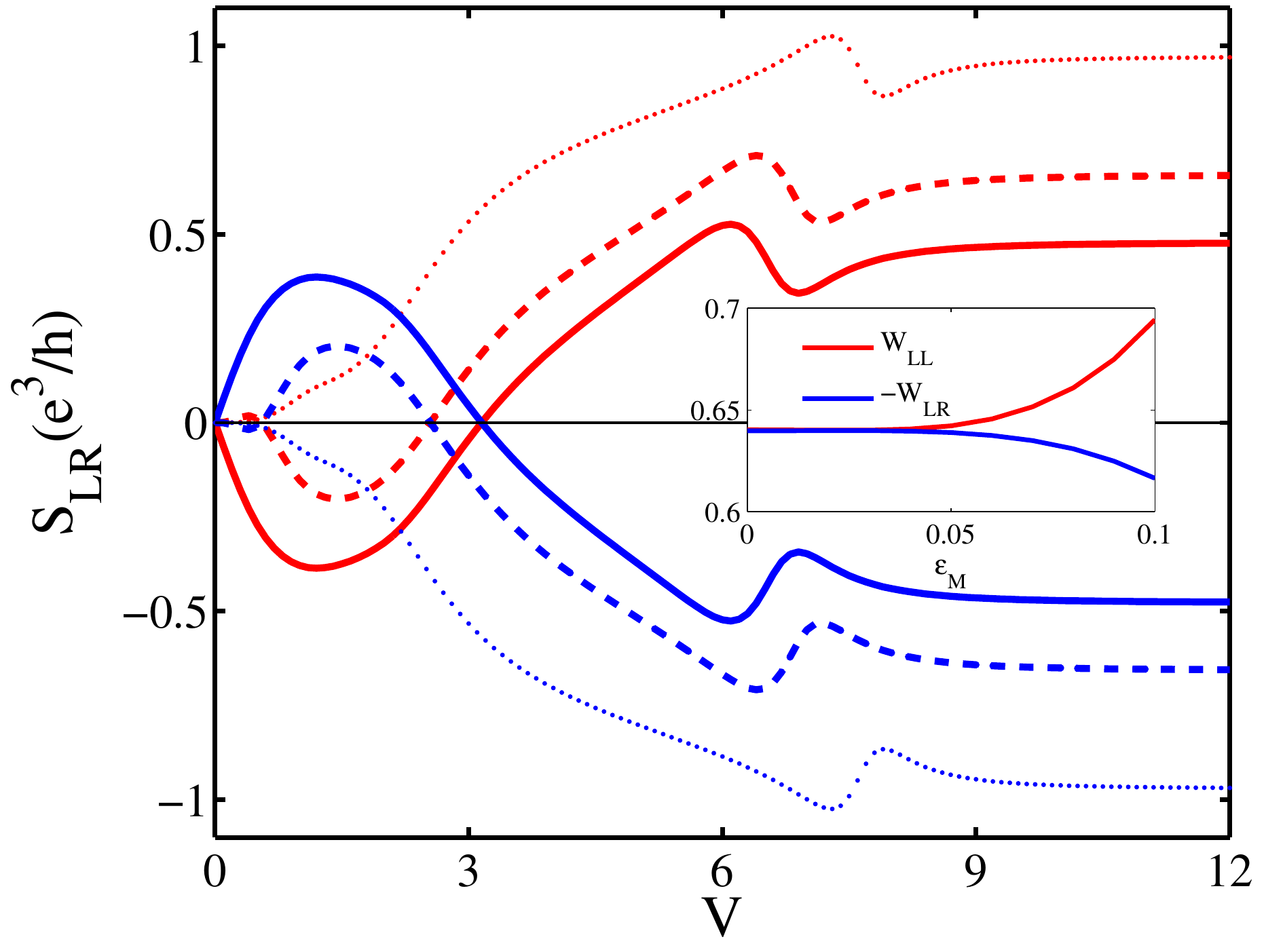}
\caption{(Color online) Current cross-correlations $S_{LR}$ under SBV with $\varepsilon_L=-\varepsilon_R=2$ (red) and ASBV with $\varepsilon_L=\varepsilon_R=-2$ (blue) as a function of the voltage with different overlaps between the two MBSs as $\varepsilon_M=0$ (solid lines), $\varepsilon_M=2$ (dashed lines) and $\varepsilon_M=4$ (dotted lines). Inset: Differential current correlations defined in Eq.\,(\ref{eq22}) versus overlap strength at $V=0.01$. The QDs-MBSs couplings are $\lambda_L=4$ and $\lambda_R=2$.}  \label{figS12}
\end{center}
\end{figure}

To gain more analytical insights on the linear current correlations, we obtain in the zero-bias limit and $\varepsilon_M=0$ the differential correlations $\mathcal{W}_{\alpha\beta}=dS_{\alpha\beta}/dV$ as follows (in unit of $e^3/h$)
\begin{equation}
\mathcal{W}_{LL}(V\rightarrow0^+)=-\mathcal{W}_{LR}(V\rightarrow0^+)=P,\label{eq22}
\end{equation}
for the SBV, and
\begin{equation}
\mathcal{W}_{LL}(V\rightarrow0^+)=\mathcal{W}_{LR}(V\rightarrow0^+)=P,\label{eq23}
\end{equation}
for the ASBV. Interestingly, $P=4\mathcal{T}_{LR}^{mm}(0)=4\mathcal{T}_{LR}^{m\bar m}(0)$, which can be tested by tuning the dot levels according to eq.\,(\ref{eq11}) to extremely suppress the AR process and comparing the conductance and differential current correlations. Furthermore, the linear cross-correlation is always negative (positive) under SBV (ASBV), irrespective of the specific model parameters. For SBV, eq.\,(\ref{eq22}) indicates the linear self-correlation $S_{LL}=\mathcal{W}_{LL}\Delta V$ equals the cross-correlation $S_{LR}=\mathcal{W}_{LR}\Delta V$ except a minus sign, which is robust at small overlap [inset of fig.\,\ref{figS12}}. Considering the current conservation condition $I_L+I_R+I_{w}=0$, this feature implies the net current flowing through the nanowire is noiseless since $S_{ww}=-S_{LL}-S_{RR}-2S_{LR}=0$, which underlies the fermionic nature \cite{Freyn2010,Henny1999} of MBSs though they are based on the superconductivity. Such noiseless feature goes beyond the previous noiseless current due to the MBS-induced resonant Andreev reflection \cite{Law2009}, since Eq.\,(\ref{eq22}) is always valid regardless of the on-resonance ($T_{\alpha\beta}^{mn}=1$) or off-resonance ($T_{\alpha\beta}^{mn}\ne1$) situation. Particularly, this feature is distinct from that the difference of currents $I_L-I_R$ is noiseless in a BCS Cooper pair beam splitter with highly transmitting interfaces \cite{Freyn2010}. At the high symmetric points $\Lambda_{\pm}$, $P$ reduces to an universal value $1$ which is similar to the one reported by Liu \textit{et al.} who justified it as a unique signature of MBSs \cite{Liu2015}.

\section{Summary}\label{sum}
In conclusion, the currents and current correlations in a normal lead-double QD-TSNW hybrid beam splitter are investigated. We find the linear cross-correlations at zero temperature change signs versus the overlap between the two MBSs. Under symmetric bias voltages, the net current flowing through the nanowire is noiseless. These two features highlight the fermionic nature of such exotic Majorana excitations. Moreover, we address a unique local p-h symmetry inherited from the self-Hermitian property of MBSs, which can be revealed through measuring the currents under complementary bias voltages. All these predictions are observable in transport measurements by applying appropriate bias voltages and carefully tuning the discrete levels in QDs. Additionally, we note the nonlocal advantage of such beam pair splitter device facilitates experimentalists to exclude possible local mist in the unambiguous identification of MBSs.

\acknowledgments
Cao thanks R\'{e}gis M\'{e}lin for correspondence. This work is supported partly by NSFC, and the national program for basic research of China.

\end{document}